\documentclass[prb,aps,amsmath,showpacs,twocolumn]{revtex4}

\usepackage{graphicx}

\usepackage{dcolumn}

\usepackage{bm}

\newcommand{\ket}[1]     { | {#1} \rangle }
\newcommand{\bra}[1]     { \langle {#1} | }

\begin{document}

\title{Dynamical mean-field theory of photoemission spectra of
actinide compounds}
\author{A. Svane}
\affiliation{Department of Physics and Astronomy, 
University of Aarhus, DK-8000 Aarhus C, Denmark}
\date{\today}

\begin{abstract}
A model of photoemission spectra of actinide compounds is presented.
The complete multiplet spectrum of a single ion is calculated
by exact diagonalization of the two-body Hamiltonian of the $f^n$ shell. A coupling
to auxiliary fermion states models the interaction with a conduction sea. The
ensuing self-energy function is combined with a band Hamiltonian of the compound,
calculated in the local-density approximation, to
produce a solid state Green's function. The theory is applied to PuSe and
elemental Am. 
For PuSe a sharp resonance at the Fermi level arises from mixed valent behavior,
while several features at larger binding energies can be identified with quantum
numbers of the atomic system. For Am
the ground state is dominated by the $\ket{f^6;J=0}$ singlet but the strong
coupling to the conduction electrons mixes in a significant amount of $f^7$ character.
\end{abstract}

\maketitle

The electronic structure of actinides has been subject of extensive 
experimental\cite{puse-gouder,havela,wachter,pucoga5,ute-lander,durakiewicz}   
and theoretical\cite{Soderlind,kotliar,leon,MLM,borje,Zwicknagl,Opahle,Oppeneer,kotliar-am}   
investigations in recent years. The key issue is the 
character of the $f$-electrons which show varying degrees of band-like and/or localized
behavior. Photoemission in particular provides energy resolved information about the
$f$-electrons, the understanding of which is far from complete.\cite{Allen}
The actinide antimonides and chalcogenides\cite{puse-gouder,durakiewicz} display 
narrow band-like features around the Fermi level as well as atomic-like features
at higher binding energies. Similarly, among the elements a distinct shift of
$f$-character away from the Fermi level happens from Pu to Am,\cite{havela,naegele}
 where also the cohesive
properties suggest a shift from itinerant 
to localized
behavior of the $f$-electrons.
The atomic-like features of photoemission spectra are not 
well analyzed in terms of multiplets of isolated actinide atoms, in contrast to the
situation for rare earth compounds.\cite{Campagna}
On the other hand, 
band structure calculations generally lead to narrow $f$-bands pinned at the Fermi
level but  no structures at large binding energy, and additional modelling has been introduced to
account for the dual character of the
$f$-electrons.\cite{kotliar,leon,MLM,borje,Oppeneer,kotliar-am}
The recent development of dynamical mean-field therory (DMFT) has in particular
spawned advancements.\cite{kotliar,kotliar-am}

In the present work a novel model of actinide photoemission is presented, 
which is capable of
describing both high and low energy excitations. In applications to PuSe and Am metal
it is shown that interaction of the $f$-electrons with the conduction electrons leads to complex 
ground state configurations for the solid, which account for the non-trivial features of the
photoemission spectra. The theory is based on the Hubbard-I approach suggested in
Ref. \onlinecite{LDA++} but augmented with coupling to the conduction sea.

To model the photoemission of actinide compounds 
the $\mathbf{k}$-integrated spectral function $A(\omega)=\pi^{-1}\text{Im} G^{loc}
(\omega)$, 
with
\begin{equation}
         \label{spectral}
  G^{loc}(\omega)  =  
                        \frac{1}{N_k} \sum_\mathbf{k} G_\mathbf{k} (\omega) 
\end{equation}
is calculated. 
A dynamical mean-field,
$\Sigma(\omega)$, is added to a band structure Hamiltonian, which 
is calculated by standard density-functional-theory based methods,\cite{OKA}
and the crystal Green function $G_\mathbf{k}$ 
obtained by inversion:
\begin{equation}
         \label{crysgreen}
G_\mathbf{k} (\omega)  =  \left(\omega - \Sigma (\omega) - 
H^{\text{LDA}}_ \mathbf{k}\right)^{-1}.
 \end{equation}
$\Sigma(\omega)$
is calculated from an effective impurity model 
describing an isolated atom coupled to a bath of uncorrelated electrons:
\begin{equation}
         \label{Himp}
\hat{H}_{imp}=\hat{H}_{atom}+ \hat{H}_{coup}.
\end{equation}
The atomic Hamiltonian, $\hat{H}_{atom}$, 
includes the electron-electron interaction, $\hat{V}_{ee}$,
 within the $f^n$ shell of an isolated actinide atom: 
  \begin{eqnarray}
  \label{eqHatom}  
   \hat{H}_{atom} & = &
(\epsilon_f-\mu)\sum_m f_{m}^{+} f_{m}
  +  \xi \sum_{i} \vec{s}_{i} \cdot \vec{l}_{i}
\nonumber \\
 & + & 
 \frac{1}{2}\sum_{m_j} U_{m_1m_2m_3m_4} f_{m_1}^{+}
 f_{m_2}^{+} f_{m_3}  f_{m_4} 
 \end{eqnarray}
Here, indices $m_i=1,..,14$ refer to the individual orbitals  of the
$f^n$ shell including spin, and $f_{m}^{+}$ and $f_{m}$ are the creation and annihililation
operators. The bare $f$-level is 
$\epsilon_f$,
and $\mu$ is the chemical potential.
The second term is 
the spin-orbit energy, $\xi$ being the spin-orbit constant,
which is calculated from the self-consistent band structure potential, and 
$\vec{s}_{i}$ and $\vec{l}_{i}$ are the spin and orbital moment operators for the
$i$'th electron.

The two-electron integrals of the Coulomb operator, 
$U_{m_1m_2m_3m_4}=\langle m_1m_2|\hat{V}_{ee}|m_3m_4\rangle
$,
may be expressed in terms of Slater integrals, $F^k$, $k=0,2,4,6$,\cite{LDA++}
which are computed with the 
self-consistent radial $f$-waves of the band structure calculation.
In practice we find it necessary to reduce the leading Coulomb integral
$F^0=U_{m_1m_2m_1m_2}$ from its bare value, similar to other studies.\cite{OG}
The physical reason for this
is the significant screening of the charge-fluctuations on the
$f$-shell by the fast conduction electrons, which happen as part of the photoemission
process, but which is not treated in the present model. A complete treatment would
involve the computation of the solid dielectric function, as is done in the
GW-approximation.\cite{Hedin,Silke} The higher
Slater integrals, $F^2$, $F^4$ and $F^6$, govern the energetics of orbital fluctuations
within the $f^n$ shell, for which the screening is less important, and we therefore use
the {\it ab-initio} calculated values for those parameters.

The coupling to a sea of conduction electrons
is modelled as a simple hopping term for each of
the 14 $f$-electrons into an auxiliary state at the Fermi level:
  \begin{eqnarray}
  \label{Hcoup}  
   H_{coup} =\mu \sum_m c_{m}^{+}c_{m}+V\sum_m\left(c_{m}^{+}f_{m}+
                          f_{m}^{+}c_{m} \right).
 \end{eqnarray}
The impurity Hamiltonian,  Eq. (\ref{Himp}), is solved by exact diagonalization, and
the dynamical mean-field, 
$\Sigma_{mm^{\prime}}$, follows from the 
impurity Green's function,
$(G^{imp})_{mm^{\prime}}$ ,
which is computed from the eigenvalues, $E_{\mu}$, and eigenstates, $\ket{\mu}$, of
$\hat{H}_{imp}$ as:
  \begin{eqnarray}
         \label{eqGimp}
         (G^{imp})_{mm^{\prime}} (\omega)  =  \sum_{\mu\nu} w_{\mu\nu}
         \frac{ \bra{\mu}f_{m}\ket{\nu}  \bra{\nu}f^{+}_{m^{\prime}}\ket{\mu} }
          {\omega+E_{\mu}-E_{\nu}} \\
         \Sigma_{mm^{\prime}} (\omega)  =  
            (G^{0})^{-1}_{mm^{\prime}} (\omega)
         -  (G^{imp})^{-1}_{mm^{\prime}} (\omega).
 \end{eqnarray}
In the zero-temperature limit, $w_{\mu\nu}=1$ if either state $\ket{\mu}$ or
$\ket{\nu}$ is the ground state, and $w_{\mu\nu}=0$ otherwise.
Further, $G^{0}=(\omega-\epsilon_f+\mu)^{-1}$ is the Green's function of the bare 
$f$-level. Alternatively, $G^{0}$ may be determined by the DMFT self-consistency
through $(G^{0})^{-1}=(G^{loc})^{-1}+\Sigma(\omega)$, whith $G^{loc}$ given by Eq. (1).
In the present applications there are only insignificant difference
between these two expressions for $G^{0}$, {\it i.e.} the DMFT self-consistent
solid state local Green's function
is fairly close to that of a bare f-electron (at the energy position given by the LDA
eigenvalue).
%
The effect of the coupling term in Eq. (\ref{Hcoup})
is to mix the multiplet eigenstates corresponding
to fluctuations in $f$-occupancy. The eigenstates of $\hat{H}_{imp}$ will not contain an integral number
of $f$-electrons, but may deviate more or less from the ideal atomic limit, depending on
parameters, in particular the coupling strength $V$.

In a more realistic treatment the conduction states in (\ref{Hcoup})  should aquire a width, and
the hybridization parameter $V$ be energy dependent.\cite{OG} The 
energy dependent hybridization may be calculated directly
from a LDA band structure calculation (or from the self-consistent DMFT bath Green's
function),\cite{OG} but the straightforward mapping to the above model leads to overestimation
of the interaction, since the auxiliary  electrons cost  no excitation energy.
Hence, $V$ is treated as an
adjustable parameter of the present theory.
The hopping parameter $V$ in (\ref{Hcoup}) could in principle depend on orbital index
(which would be relevant in highly anisotropic crystal structures\cite{Zwicknagl}), but
here we apply the isotropic model of Eq. (\ref{Hcoup}). 
Similarly, with no particular extra effort the bare $f$-energies 
in (\ref{eqHatom}) could be orbital dependent, e.g.
for studies of crystal field effects. In the present study only a single $f$-energy is
used, given by its average value in the band structure calculation.


\begin{figure}
\begin{center}
\includegraphics[width=90mm,clip]{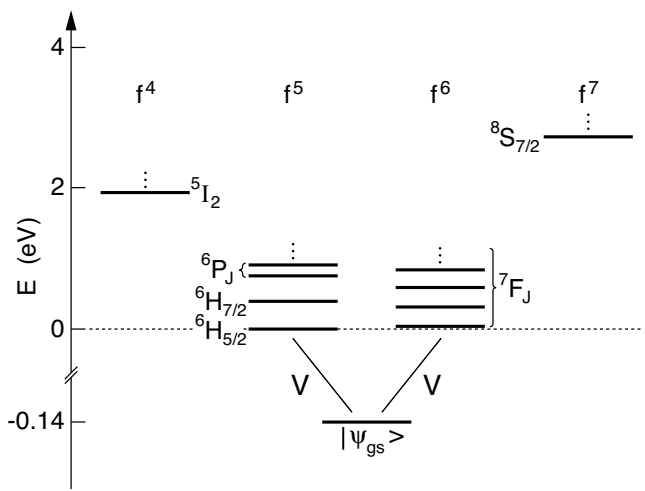}\\
\caption{Illustration of the lowest multiplets for a Pu atom at parameter values
pertinent to PuSe [\onlinecite{Puparm}]. The energy is reckoned from the lowest
$\ket{f^5;J=5/2}$ state.  Only the lowest few $f^n$ states are shown.
Each level is labeled by its (approximate) Russell-Saunders term.
The ground state, when the coupling to the conduction states is switched on, is 
formed primarily as a linear combination of the lowest lying $f^5$ and $f^6$ states.
The energy gain due to the coupling is 0.14 eV; for clarity
 the effect is exaggerated on the figure.
}
\end{center}
\label{fig1}
\end{figure}
The theory outlined above may be applied to any $f$-element, {\it i.e.}
any $f$-occupancy. The chemical potential
$\mu$  determines the ground state of the model. Thus, this
parameter must be adjusted to match the $f$-element under study. The adjustment of $\mu$
to a certain extent balances any inaccuracies in the values of the primary Slater
integral, $F^0$, since a shift in the latter may be compensated by a shift
in $\mu$. 
With $F^0$ and $\mu$ fixed, the energies of the ground and excited state multiplets 
of $\hat{H}_{atom}$,
$\ket{f^n;JM}$ are determined.  Fig. 1 ilustrates this for parameters relevant for
a Pu atom in PuSe.\cite{Puparm,RS} This selects
$\ket{f^5;J=5/2}$  as the ground state with $\ket{f^6;J=0}$ at marginally 44 meV
higher energy.  

Introducing the coupling to the auxiliary states,
Eq. (\ref{Hcoup}), leads to a mixed ground state,
composed of 70 \% $f^5$ and 30 \% $f^6$, with 67 \% and 24 \% stemming from the
lowest $\ket{f^5;J=5/2}$ and $\ket{f^6;J=0}$ states, respectively. Similar mixing
occurs in all excited states, and all the degeneracies of the atomic multiplets are
completely lifted by the coupling. In particular, the ground state is non-magnetic,
and the local-moment character of
the $f^5$ ion will only commence at elevated temperatures.
A number of eigenstates occur at low excitation energies and allow for low energy
transitions in the photoemission. These transitions would be absent in a pure
$\ket{f^5;J=5/2}$ ground state.

\begin{figure}
\begin{center}
\includegraphics[width=90mm,clip]{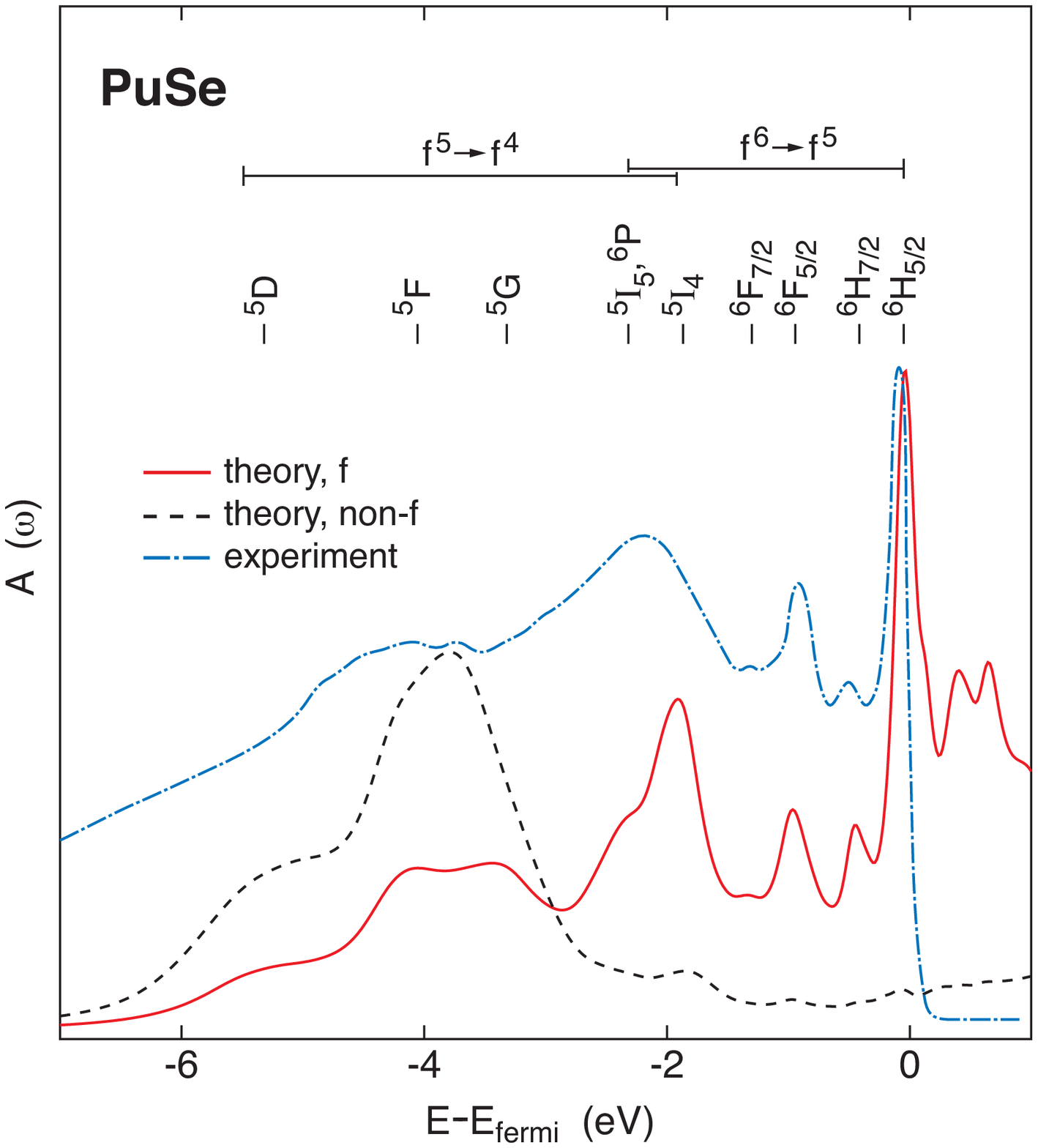}\\
\caption{(Color online) 
Calculated spectral function for PuSe, $f$-part in full line (red) and non-$f$ in dashed
line, compared to the experimental photoemission spectrum of Ref.
\onlinecite{puse-gouder} (in dash-dotted and blue) at 40.8 eV photon   energy.
The energy is measured relative to the Fermi level, [\onlinecite{Broadn}].
The dominating final state atomic term is indicated above each major peak.
}
\end{center}
\label{fig-puse}
\end{figure}

\begin{figure}
\begin{center}
\includegraphics[width=90mm,clip]{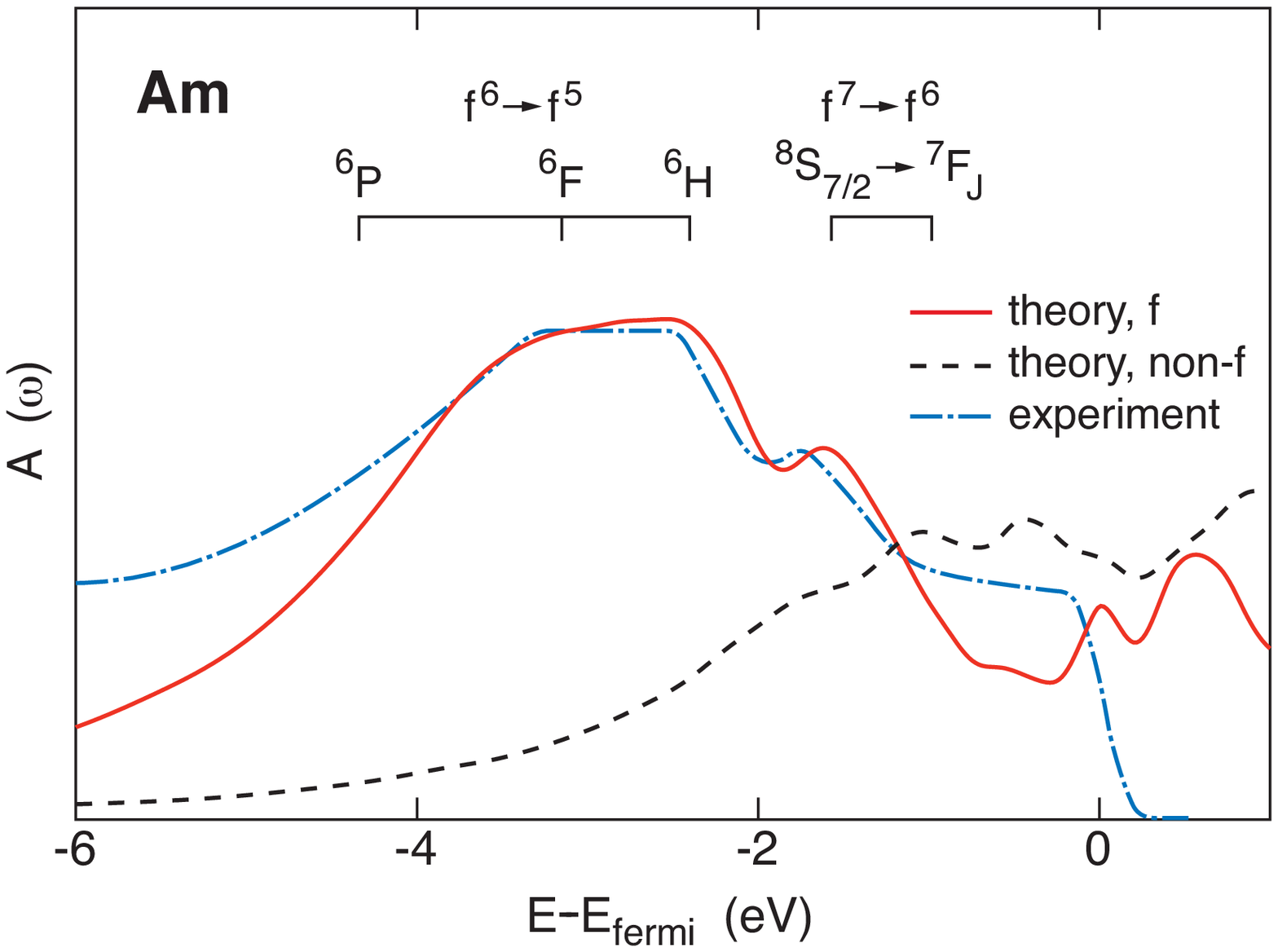}\\
\caption{(Color online) Calculated spectral function for americium metal, 
$f$-part in full line (red) and non-$f$ in dashed
line, compared to the experimental photoemission spectrum of Ref.
\onlinecite{naegele} (in dash-dotted and blue) recorded at 40.8 eV photon   energy.
The energy is measured relative to the Fermi level, [\onlinecite{Broadn}].
The dominating final state atomic term is indicated.
}
\end{center}
\label{fig-am}
\end{figure}

 The calculated spectral function, Eq. (\ref{spectral}), of PuSe is depicted in
Figure 2 and compared to the experimental photoemission spectrum of Ref.
\onlinecite{puse-gouder}. There is a one-to-one correspondence between the measured
and
calculated spectral features, and the theory allows identification of each of these. 
In particular, a sharp resonance at the Fermi level is due to the low-energy
$^7$F$_{0}\rightarrow ^6$H$_{5/2}$ excitations, which may be identified with
heavy-fermion type behavior.\cite{wachter} At 0.4 eV binding energy excitations into the
$^6$H$_{7/2}$ final state give rise to a small peak, 
while the more
pronounced peak at $\sim 1$ eV binding energy marks the excitations into the
$^6$F$_{5/2}$ final state, which also has a spin-orbit satellite, visible around
1.4 eV binding energy. 
The major emission peak around 2 eV binding energies is due to 
$^6$H$_{5/2}\rightarrow ^5$I$_{4}$ transitions with small
$^7$F$_{0}\rightarrow ^6$P$_{5/2}$ contribution.
At even larger binding energy
a broad emission coincides with the Se $p$-band, (dashed line in Fig. 2).
It is unclear whether the broad shoulder seen in the experiment between 4 and 5
eV includes emission related to $f$-features or is solely due to the Se $p$-band.
The identification of PuSe as a mixed-valent compound was made by
Wachter,\cite{wachter} and the present work supports this view. Details of the
identifications of the spectral features are different, though, in particular 
in the present theory it is not necessary to reduce the multiplet splittings
within each $f^n$ configuration compared to the atomic values.
The matrix elements in the numerator of Eq. (6) ensures that all atomic selection rules are
obeyed,
and determine the weights of the individual peaks. 
Left out of the present treatment is the matrix
element between the $f$-electron one-particle wave and the the photoelectron, which may be
assumed to vary only slowly with energy over the range of the valence bands. 
However, the experimental spectrum contains contributions from all allowed transitions,
$f$- as well as non-$f$-related. 
The justification for the comparison in Fig. 2 (and in Fig. 3 for americium below) 
of the theoretical  $f$-spectral function with the experimental photoemission spectrum
recorded with 40.8 eV photons is that the cross section strongly favors 
emission of an $f$-electron at this energy. The experiments reported in 
Ref. \onlinecite{puse-gouder} 
include photoemission spectra taken with 21.2 eV electrons, at
which energy the contribution of the non-$f$ electrons dominates, with a spectrum
quite close to the non-$f$ spectral function shown in Fig. 2.
Finally, life-time effects and secondary electron 
processes are not considered in the present theory.


\begin{figure}
\begin{center}
\includegraphics[width=90mm,clip]{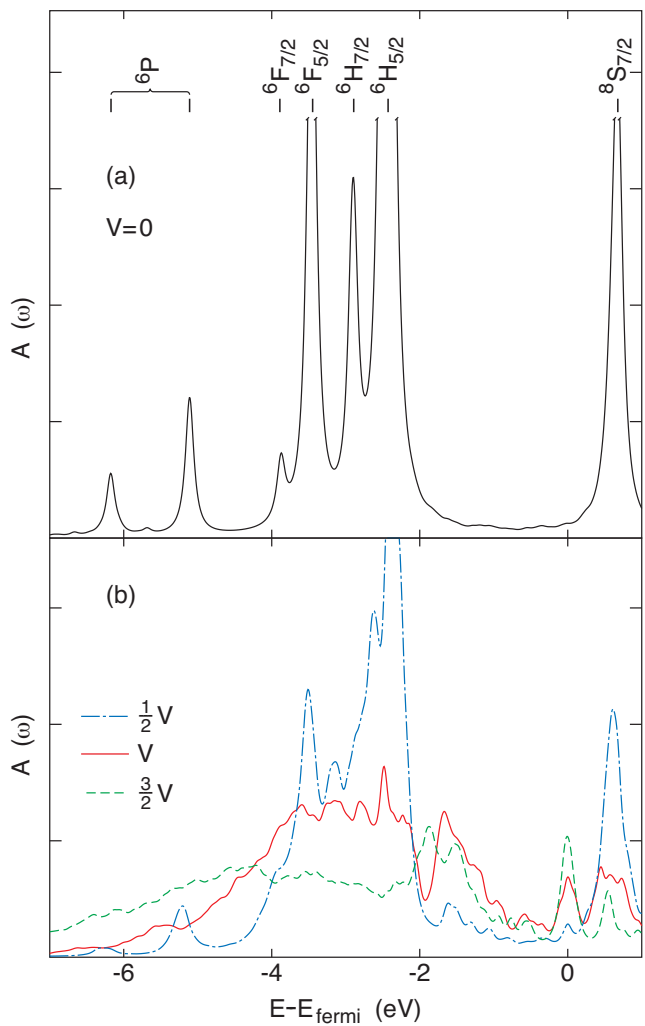}\\
\caption{(Color online)
Calculated $f$-part of the spectral function for americium metal, 
for varying strength of the coupling to the conduction electrons,
a) $V=0.0$ eV , and
b) $V=0.16$ eV (dash-dotted and blue), $V=0.33$ eV (full line and
red), and $V=0.49$ eV (dashed and green).  
The spectrum presented in Figure 3 corresponds to the full line curve of the present
plot with a larger broadening [\onlinecite{Broadn}].
}
\end{center}
\label{fig-amv}
\end{figure}

The calculated spectral function\cite{Amparm} of Am is compared to photoemission
experiment\cite{naegele} in figure 3. 
The multiplet level diagram of the isolated Am atom differs from  that of Pu shown in
Fig. 1 by shifting the balance such that $\ket{f^6;J=0}$ is now the ground state
with the $\ket{f^7; {}^8S_{7/2}}$  at 0.7 eV higher energy,
and      $\ket{f^5; {}^6H_{5/2}}$  at 2.3 eV higher energy.
Spin-orbit interaction seriously violates the Russell-Saunders coupling scheme.
 The overlap of the americium ground state and the Russell-Saunders 
$|f^6;\,{}^7F_0\rangle$ state is 0.69. The influence on the photoemission transition
probabilities is substantial. For   example,  the low-energy transitions from
the $\ket{f^6;J=0}$ ground state to the $\ket{f^5;{}^6H_{5/2}}$ and
$\ket{f^5;{}^6H_{7/2}}$  occur in ratio $2.6:1$, while Russell-Saunders coupling 
predicts the ratio $1:2.5$,\cite{Campagna} 
i.e. exactly reversed weight on the two components.
For the same reason, the $J=7/2$ satellite of the peak at the Fermi level in PuSe,
Fig. 1, is significantly weaker compared to the main peak.

The 
coupling to the conduction electrons, Eq. (\ref{Hcoup}), leads to further 
discrepancy with the free atom Russell-Saunders ground state. 
Good     agreement
with the experimental spectrum\cite{naegele} is obtained with a coupling
parameter $V=0.33$ eV in Eq. (\ref{Hcoup}). The ensuing Am ground state remains a
singlet but attains a significant (16 \%) admixture of $\ket{f^7; {}^8S_{7/2}}$ character.
The $\ket{f^6;J=0}$ state has still got 61 \% weight in the coupled ground state,
while the remaining weight scatters over the higher levels of the $f^6$
configuration. The effect on the spectral function is drastic, in particular by
leading to significant emission of $\ket{f^7; {}^8S_{7/2}}\rightarrow \ket{f^6;J}$
peaking at      1.7 eV binding energy with a broad tail towards lower binding
energy, i.e. this transmission accounts for the shoulder observed in the experiment
around 1.8 eV. The main emission of $\ket{f^6; {}^7F_{0}}\rightarrow 
\ket{f^5; {}^6H, {}^6F, {}^6P}$ occurs from $\sim 2.5$ eV and higher binding energies, and accounts 
for the plateau seen in both experiment and theory between 2.5 and 3.5 eV.

In Figure 4 the evolution of the calculated Am spectrum with the 
strength of the coupling
to the conduction electrons is shown.
In Fig. 4(a) the coupling has been set to zero, and the spectrum shows the dominating
emission between 2 and 4 eV binding energy due to 
$\ket{f^5; {}^6H_{5/2}}$ and $\ket{f^5; {}^6F_{5/2}}$ final states and their $J=7/2$
satellites. The small peaks at -5.1 and -6.2 eV are remnants of the 
emission to $\ket{f^5; {}^6P_{5/2}}$ final states, highly distorted due to spin-orbit
interaction in both initial and final state. 
In Fig. 4(b) the distortion of the spectral function due to coupling to the
conduction sea is illustrated, at $V=\frac{1}{2}$, $1$ and $\frac{3}{2}$ times the value used in
Figure 3. At binding energies less than 1.7 eV the triangular structure evolves due
to $\ket{f^7; {}^8S_{7/2}}$ admixture into the initial state. At the highest value of
$V$ the emission broadens considerably. At the Fermi level a peak grows up with
increasing coupling strength, visible as a small shoulder in Figure 3, but not
resolved in the experimental data.


In conclusion, a theory has been developed which allows a detailed description of
the photoemission spectra of actinide compounds. The three most important energy
scales are the $f$ intra-shell Coulomb interactions, the spin-orbit interaction, and
the coupling to the conduction electrons, which are all incorporated in the theory.
All atomic selection rules for the photoexcitation are obeyed.
The key approximation is that of a dynamical mean-field calculated for a single
actinide ion with a simplified interaction with conduction electrons.
Compared to recent advancements in the field\cite{kotliar,kotliar-am} the present approach 
treats the impurity atom essentially exact including interaction with the
conduction sea.
Hence, the theory applies best to systems on the localized side of the Mott transition
of the $f$ shell. In particular, 
for PuSe and Am metal 
the interaction leads to formation of
complex ground states which strongly influence the photoemission spectra, since
the initial states of the photoemission process are not free-atom like. 
The embedment into the solid via the LDA band Hamiltonian in Eq. (\ref{crysgreen}) 
maintains and broadens
the atomic features. 
The DMFT self-consistency cycle has only minute influence on the 
calculated spectra. 


This work was partially funded by the EU Research Training Network
(contract:HPRN-CT-2002-00295) 'f-electron'.
Support from the Danish Center for Scientific Computing is acknowledged. 



\begin{thebibliography}{}
\bibitem{puse-gouder}
T. Gouder, F. Wastin, J. Rebizant, and L. Havela,
Phys. Rev. Lett. {\bf 84}, 3378 (2000).

\bibitem{havela}
L. Havela, F. Wastin, J. Rebizant, and T. Gouder,
Phys. Rev. B {\bf 68}, 85101 (2003).

\bibitem{wachter} 
P. Wachter, Sol. State Commun. {\bf 127} 599 (2003).

\bibitem{pucoga5}
J. J. Joyce, J. M. Wills, T. Durakiewicz, M. T. Butterfield, E. Guziewicz, 
J. L. Sarrao, L. A. Morales, A. J. Arko, and O. Eriksson,
Phys. Rev. Lett. {\bf 91}, 176401 (2003).

\bibitem{ute-lander}
T. Durakiewicz, C. D. Batista, J. D. Thompson, C. G. Olson,  J. J. Joyce, 
G. H. Lander, J. E. Gubernatis, M. T. Butterfield, A. J. Arko, J. Bonca, 
K. Mattenberger, and O. Vogt,
Phys. Rev. Lett. {\bf 93}, 267205 (2004).

\bibitem{durakiewicz}
T. Durakiewicz, J. J. Joyce, G. H. Lander, C. G. Olson, M. T. Butterfield, 
E. Guziewicz, A. J. Arko, L. Morales, J. Rebizant, K. Mattenberger, and O. Vogt,
Phys. Rev. B {\bf 70}, 205103 (2004).


\bibitem{Soderlind} P. S\"oderlind,
Adv. Phys.       {\bf 47}, 959 (1998).

\bibitem{kotliar}
S. Y. Savrasov, G. Kotliar, and E. Abrahams,
Nature, {\bf 410}, 793 (2001).
%

\bibitem{leon} L. Petit, A. Svane, Z. Szotek and W.M. Temmerman,
Science {\bf 301}, 498 (2003).

\bibitem{MLM} J. M. Wills, O. Eriksson, A. Delin, P. H. Andersson, J. J. Joyce,
T. Durakiewicz, M. T. Butterfield, A. J. Arko, D. P. Moore, and L. A. Morales,
J. Electr. Spectr. and Rel. Phenom., {\bf 135}, 163 (2004).

\bibitem{borje} P. A. Korzhavyi, L. Vitos, D. A. Andersson, and B. Johansson,
Nature Mater. {\bf 3}, 225 (2004).

\bibitem{Zwicknagl}
D. V. Efremov, N. Hasselmann, E. Runge, P. Fulde, and G. Zwicknagl,
Phys. Rev. B {\bf 69}, 115114 (2004).

\bibitem{Opahle}   
I. Opahle and P. M. Oppeneer,
Phys. Rev. Lett. {\bf 90}, 157001 (2003);
I. Opahle, S. Elgazzar, K. Koepernik, and P. M. Oppeneer,
Phys. Rev. B {\bf 70}, 104504 (2004).

\bibitem{Oppeneer}   
A. B. Shick, V. Jani\v{s}, and P. M. Oppeneer,
Phys. Rev. Lett. {\bf 94}, 16401 (2005).

\bibitem{kotliar-am}
S. Y. Savrasov, K. Haule, and G. Kotliar,
Phys. Rev. Lett. {\bf 96}, 036404 (2006).

\bibitem{Allen}  
J. W. Allen, Y.-X. Zhang, L. H. Tjeng, L. E. Cox, M. B. Mable, and C.-T. Chen,
J. Electr. Spectr. and Rel. Phenom., {\bf 78}, 57 (1996).

\bibitem{naegele}
J. R. Naegele, L. Manes, J. C. Spirlet, and W. M\"uller,
Phys. Rev. Lett. {\bf 52}, 1834 (1984).

\bibitem{Campagna}
M. Campagna, G. K. Wertheim and  Y. Baer,
in {\it Photoemission in Solids II}, Eds. L. Ley and M. Cardona,
(Springer, Berlin, 1979), ch. 4.

\bibitem{LDA++}
A. I. Lichtenstein and M. I. Katsnelson,
Phys. Rev. B {\bf 57}, 6884 (1998).

\bibitem{OKA} We use the tight-binding linear muffin-tin-orbital method, see
O. K. Andersen, Phys. Rev. B {\bf 12}, 3060 (1975);
O. K. Andersen and O. Jepsen, Phys. Rev. Lett. {\bf 53}, 2571 (1984).

\bibitem{OG}
O. Gunnarsson, O. K. Andersen, O. Jepsen, and J. Zaanen,
Phys. Rev. B. {\bf 39}, 1708 (1989).

\bibitem{Hedin}
L. Hedin, Phys. Rev. {\bf 139}, A796 (1965).

\bibitem{Silke}
S. Biermann, F. Aryasetiawan, and A. Georges,
Phys. Rev. Lett. {\bf 90}, 86402 (2003); 
F. Aryasetiawan, M. Imada, A. Georges,
G. Kotliar, S. Biermann, and A. I. Lichtenstein, 
Phys. Rev. B. {\bf 70}, 195104 (2004).

\bibitem{Puparm} 
For a Pu atom in PuSe we calculate: $F^2=7.9$ eV, $F^4=5.0$ eV, $F^6=3.6$ eV,
$\xi=0.28$ eV. The screened Coulomb parameter is taken as $F^0=2.7$ eV, and the
chemical potential fixed at $\epsilon_f-\mu=-10.8$ eV. The coupling strength
used in Fig. 2 is $V=0.12$ eV. The crystal structure is NaCl with the experimental
lattice constant $a=5.79$ \AA.


\bibitem{RS} 
For the sake of presentation the approximate term values for the
eigenstates of $\hat{H}_{atom}$ are used throughout, even 
though $L$ and $S$ are not good quantum numbers.


\bibitem{Broadn} 
Experimental broadening is simulated by evaluating 
the spectral functions in Figs. 2, 3 and 4
at complex energies, given by $\omega=(E-E_{fermi})\cdot(1+ib)$,
where $b=0.05$ in Fig. 2 and
$b=0.075$ in Fig. 3, while $\omega=(E-E_{fermi})+i\delta$, with
$\delta=65$ meV is used   in Fig. 4. 

\bibitem{Amparm} 
For an Am atom in americium metal we calculate: $F^2=8.7$ eV, $F^4=5.6$ eV, $F^6=4.0$ eV,
$\xi=0.34$ eV. The screened Coulomb parameter is taken as $F^0=3.0$ eV, and the
chemical potential fixed at $\epsilon_f-\mu=-14.3$ eV. The coupling strength
used in Fig. 3 is $V=0.33$ eV. The crystal structure was taken to be fcc
with the experimental equilibrium volume.



\end{thebibliography}
\end{document}